\documentclass[manuscript,nonacm]{acmart}
\renewcommand\footnotetextcopyrightpermission[1]{} 
\AtBeginDocument{
  }

\usepackage{threeparttable}
\usepackage{fancyhdr}
\usepackage{graphicx}
\usepackage{xcolor} 
\usepackage{url}
\usepackage{subcaption}
\usepackage{booktabs}
\usepackage{tabularx}
\usepackage{caption}
\usepackage{hyperref}
\usepackage{enumitem}
\usepackage{soul}
\usepackage[inkscapelatex=false]{svg}
\usepackage{xpatch}

\makeatletter
\let\@authorsaddresses\@empty
\makeatother

\begin{document}
\title{Algorithmic Mirror: Designing an Interactive Tool to Promote Self-Reflection for YouTube Recommendations}
\begingroup
\renewcommand\thefootnote{}\footnote{
 This paper was presented at the 2025 ACM Workshop on Human-AI Interaction for Augmented Reasoning (AIREASONING-2025-01). This is the authors’ version for arXiv.}
\endgroup

\author{Yui Kondo}
\affiliation{%
  \institution{Oxford Internet Institute, University of Oxford}
  \country{UK}}

\author{Kevin Dunnell}
\affiliation{%
  \institution{MIT Media Lab, Massachusetts Institute of Technology}
  \city{Cambridge}
  \country{USA}}

\author{Qing Xiao}
\affiliation{%
 \institution{Human-Computer Interaction
Institute, Carnegie Mellon University}
 \state{Pennsylvania}
 \country{USA}}

\author{Jun Zhao}
\affiliation{%
  \institution{Department of Computer Science, University of Oxford}
  \country{UK}
}

\author{Luc Rocher}
\affiliation{%
  \institution{Oxford Internet Institute, University of Oxford}
  \country{UK}}

\begin{abstract}

Big Data analytics and Artificial Intelligence systems derive non-intuitive and often unverifiable inferences about individuals' behaviors, preferences, and private lives. Drawing on diverse, feature-rich datasets of unpredictable value, these systems erode the intuitive connection between our actions and how we are perceived, diminishing control over our digital identities. While Explainable Artificial Intelligence scholars have attempted to explain the inner workings of algorithms, their visualizations frequently overwhelm end-users with complexity. This research introduces ‘hypothetical inference’---a novel approach that uses language models to simulate how algorithms might interpret users' digital footprints and infer personal characteristics without requiring access to proprietary platform algorithms. Through empirical studies with fourteen
adult participants, we identified three key design opportunities to foster critical algorithmic literacy: (1) reassembling scattered digital
footprints into a unified map, (2) simulating algorithmic inference
through LLM-generated interpretations, and (3) incorporating temporal dimensions to visualize evolving patterns. This research lays the groundwork
for tools that can help users recognize the influence of data on platforms and develop greater autonomy in increasingly
algorithm-mediated digital environments.
  
\end{abstract}

\keywords{Critical Algorithmic Literacy, Datafication, Recommendation Algorithm Visualization, YouTube} 

\begin{teaserfigure}
  \includegraphics[width=\textwidth]{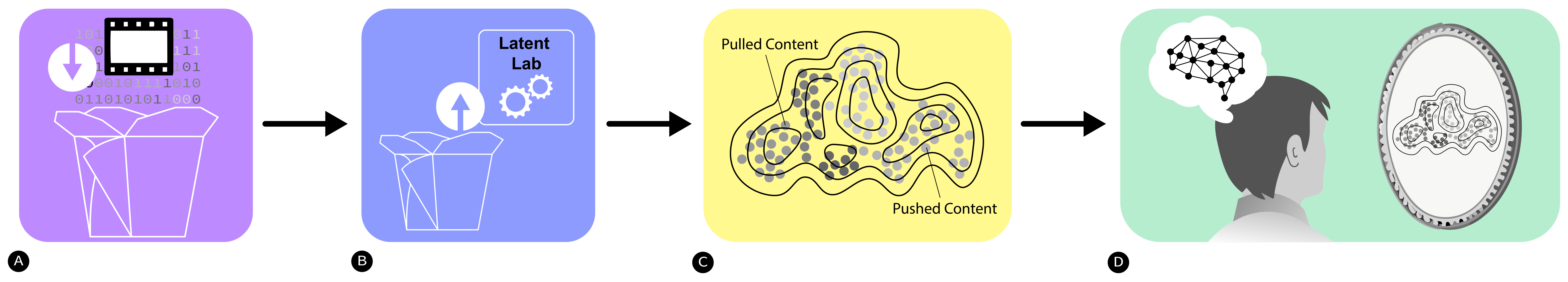}
      \caption{Four-step workflow for reflective exploration of YouTube recommendations in Algorithmic Mirror: (A) Export watch history via Google Takeout, (B) Process via semantic analysis using Latent Lab, (C) Generate a semantic map of video topics and sources (pulled vs. pushed content), and (D) Reflect on interests and algorithmic recommendations---illustrated by interconnected nodes.}
  \Description{This figure outlines a four-step workflow for fostering reflective engagement with personalized YouTube recommendations. In step (A), a user exports their YouTube viewing history using Google Takeout. In step (B), the extracted data is uploaded into a processing platform (Latent Lab System) that uses semantic analysis on the viewing records. Step (C) reveals the output: a semantic map clustering and organizing content by thematic relationships. Finally, in step (D), a user is shown facing a mirror-like interface that displays this semantic map. Above the user's head in a thought bubble, a network of interconnected nodes symbolizes the artificial intelligence algorithms that shape recommendations, now visually exposed and reflected back to the user.}
  \label{fig:teaser}
\end{teaserfigure}

\maketitle

\section{Introduction}

Online, recognizing and identifying one's own image is a difficult but important task. Recognizing oneself helps develop self-consciousness and introspection, a process that Lacan calls ‘mirror stage’~\cite{lacanFourFundamentalConcepts1977}. The mirror allows individuals to encounter their reflection and construct an image of a unified self. Online, however, our sense of self remains largely invisible to us \cite{beerDataGaze_Capitalism2019}. As individuals scroll on algorithmic feeds, online platforms monitor and assemble a mosaic of our digital footprints (e.g. browsing histories, reactions, mouse movements, and clicks) scattered across multiple sites. Collected data are used to analyze or predict aspects of an individual's life, including work performance, economic situation, health, personal preferences, interests, reliability, behavior, and location~\cite{pasqualeBlackBoxSociety2015}. 

To ensure humans' right to understand and contest AI-driven decisions, HCI and Explainable AI (XAI) scholars have prioritized algorithmic transparency. Traditional algorithm visualizations often overwhelm end-users with complexity~\cite{heInteractiveRecommenderSystems2016}, while proprietary algorithms remain protected by trade secrets~\cite{kitchinThinkingCriticallyResearching2017}. Recent personalized XAI approaches aim to make explanations more accessible and relevant to users' individual contexts, moving beyond one-size-fits-all solutions ~\cite{conatiPersonalizedHumanCenteredExplainable}. Systems like CHAITok and The Algorithm have attempted to simulate recommendation patterns using small user-selected datasets~\cite{wangCHAITokProofofConceptSystem2024, ailaelkattanAlgorithm2024}, but these simulations lack the authenticity of users' actual digital footprints. 

Meanwhile, Large Language Models (LLMs) offer promising capabilities for XAI ~\cite{wuUsableXAI102024}.  For example, some work has investigated using LLMs to directly explain ML models ~\cite{bhattacharjeeLLMguidedCausalExplainability2024,kroegerInContextExplainersHarnessing2024}, to understand user questions to generate the appropriate explanation ~\cite{watkinsACEActionControla} or to transform explanations into natural language narratives ~\cite{zytekLLMsXAIFuture}. We believe that LLMs can be used to generate plausible inferences from a user's digital footprint, potentially bridging the gap between complex algorithmic systems and intuitive user understanding.

In this study, we provide users with a personalized data visualization that we call ``Algorithmic Mirror'' for real-world algorithmic investigations. Algorithmic Mirror acts as a digital mirror, bringing together fragmented digital footprints from online platforms and time periods to reveal how platforms and algorithms shape users' inner lives over time. Our key design concepts are:
\begin{enumerate}[nosep]
\item \textit{Interactive Map Visualization}: creating a comprehensive visual representation where scattered digital footprints are reassembled into a single, navigable 2D map that displays content as individual dots clustered by semantic similarity
\item \textit{Comparison Across Platforms}: overlaying data from multiple platforms (YouTube, Spotify, etc.) to reveal distinct platform-specific behavioral patterns and enable users to visualize how their digital identities fragment across different services
\item \textit{Hypothetical Inference}: simulating how algorithms might interpret user digital footprint and infer personal characteristics generated by Large Language Models (LLMs) without access to the actual proprietary algorithms used by social media platforms
\item \textit{Temporal Dimension}: utilizing long-term data to enable users to reflect on the effects of recommendation algorithms and their actions over time
\end{enumerate}

We then examine how this design can support users' reflection on their online behavior and data, with the following research question:
\begin{itemize}[nosep]
    \item[\textbf{RQ:}]  How might a tool for simulating algorithmic inference be used to facilitate critical reflection over data? 
\end{itemize}
Taken together, our findings contribute to the research on XAI for enhancing users' critical algorithmic literacy, as well as the ability to understand and evaluate how algorithms work and impact information we see and decisions we make.

\section{Related Work}

\textbf{Users Reflection of Algorithms}
While users attempt to understand the underlying logic that governs which content is shown to them on online platforms~\cite{eslamiFirstItThen2016, raderUnderstandingUserBeliefs2015}, they often remain unaware of how their digital footprints are interpreted and transformed into algorithmic identities ~\cite{cheney-lippoldNewAlgorithmicIdentity2011, wolfDIYVideosYouTube2016}. The hidden process currently in place fosters a passive approach to content consumption and self-perception, disempowering the capacity of users to explore content actively ~\cite{baughanDonEvenRemember2022}. 

To establish users' agency in algorithmic systems, researchers have developed frameworks for data literacy ~\cite{claesDefiningCriticalData2020} and AI literacy ~\cite{longWhatAILiteracy2020}. While these efforts enhance technical understanding, they often prioritize abstract knowledge over practical application in users' everyday contexts ~\cite{wangTreatMeYour2023}. On the other hand, the Critical Algorithmic Literacy (CAL) framework emphasizes reflexive understanding of algorithmic systems through personally meaningful experiences ~\cite{kafaiTheoryBiasTheory2020, wangTreatMeYour2023, wangCHAITokProofofConceptSystem2024}. CAL integrates three essential dimensions: \textit{cognitive thinking} (understanding core computational concepts and skills), \textit{situated thinking} (learning through personal contexts and social engagement), and \textit{critical thinking} (examining broader societal implications of computational systems) ~\cite{kafaiTheoryBiasTheory2020}. In this paper, we discuss how our design can enhance CAL.

\textbf{Algorithm Visualization}
HCI scholars have proposed visualization techniques to increase users' transparency and control over recommendation algorithms~\cite{tuttFDAAlgorithms2016}. One of the forms of fostering CAL is explaining how user choices affect recommendation algorithms in a controlled environment. CHAITok and The Algorithm simulated a Tiktok-like platform using 15-20 user-selected videos to explain how their behavior affects the recommendation patterns~\cite{wangCHAITokProofofConceptSystem2024, ailaelkattanAlgorithm2024}. Rather than explaining the inner workings of recommendation algorithms, some HCI scholars have also explored visualizing the outcomes of these algorithms, such as filter bubbles or echo chambers created by personalized recommendations. Nagulendra and Vassileva~\cite{nagulendraUnderstandingControllingFilter2014} design tools that enhance user awareness of personalization mechanisms and provide a sense of control over their content streams. Gao et al.~\cite{gaoBurstYourBubble2018} develop a system that encourages users to recognize and explore diverse social opinions by visually highlighting different perspectives in news articles and comments. Gillani et al.~\cite{gillaniMeMyEcho2018} create a visualization to prompt users to reflect on the politically active segments of their Twitter network, suggesting new accounts to follow to diversify their views and reduce echo chamber effects.

However, these visualization tools to promote awareness and critical evaluation of filtering algorithms use limited scope or controlled environments. Although these controlled studies provide a foundation for cognitive thinking, they may not be applicable for everyday use in online platforms~\cite{liaoQuestioningAIInforming2020}. While they only capture a snapshot of time, recommendations are dynamic~\cite{leeAlgorithmicCrystalConceptualizing2022}, learned from users’ behavior over time, and constantly changing navigational landscape. Some data visualization tools focus on explaining recommendation algorithms from actual users' choices. For example, Taste Weights demonstrates how music recommendations are generated and allows users to adjust the factors influencing those recommendations~\cite{bostandjievTasteWeightsVisualInteractive2012}. Gobo, an open-source social media browser, offers another approach by enabling users to filter and manage content from multiple platforms according to their preferences~\cite{bhargavaGoboSystemExploring2019}. 
Nevertheless, even a fully transparent `glass-box' algorithm may reduce user agency. As He et al.~\cite{heInteractiveRecommenderSystems2016} found, most algorithm visualizations are too complex for end-users to find useful and trigger information overload and frustration rather than clarity.

\textbf{Visualizing Embeddings}
Embedding visualization offers a powerful approach for real-world algorithmic investigations, revealing how algorithms interpret and profile users through their digital footprints. Embedding models transform unstructured user data—such as YouTube watch history and Spotify listening patterns—into high-dimensional representations that capture latent patterns central to recommendation systems ~\cite{GlobalGeometricFrameworkforNonlinearDimensionalityReduction}. Visualization tools like the EmbeddingProjector~\cite{smilkov2016embeddingprojectorinteractivevisualization} allow users to explore these embeddings through dimensionality reduction techniques, including linear methods like PCA~\cite{PCA} and nonlinear methods such as t-SNE~\cite{tsne} and UMAP~\cite{umap}. While these tools simplify dimensionality reduction and visualization, they often lack the ability to preprocess unstructured content into embeddings or label clusters within visualizations—critical steps for effectively applying these tools to real-world use cases~\cite{10.1145/2669557.2669559}.

\section{Design Prototype: Algorithmic Mirror}
\begin{figure*}[t]
    \centering
    \includegraphics[width=1.0\textwidth]{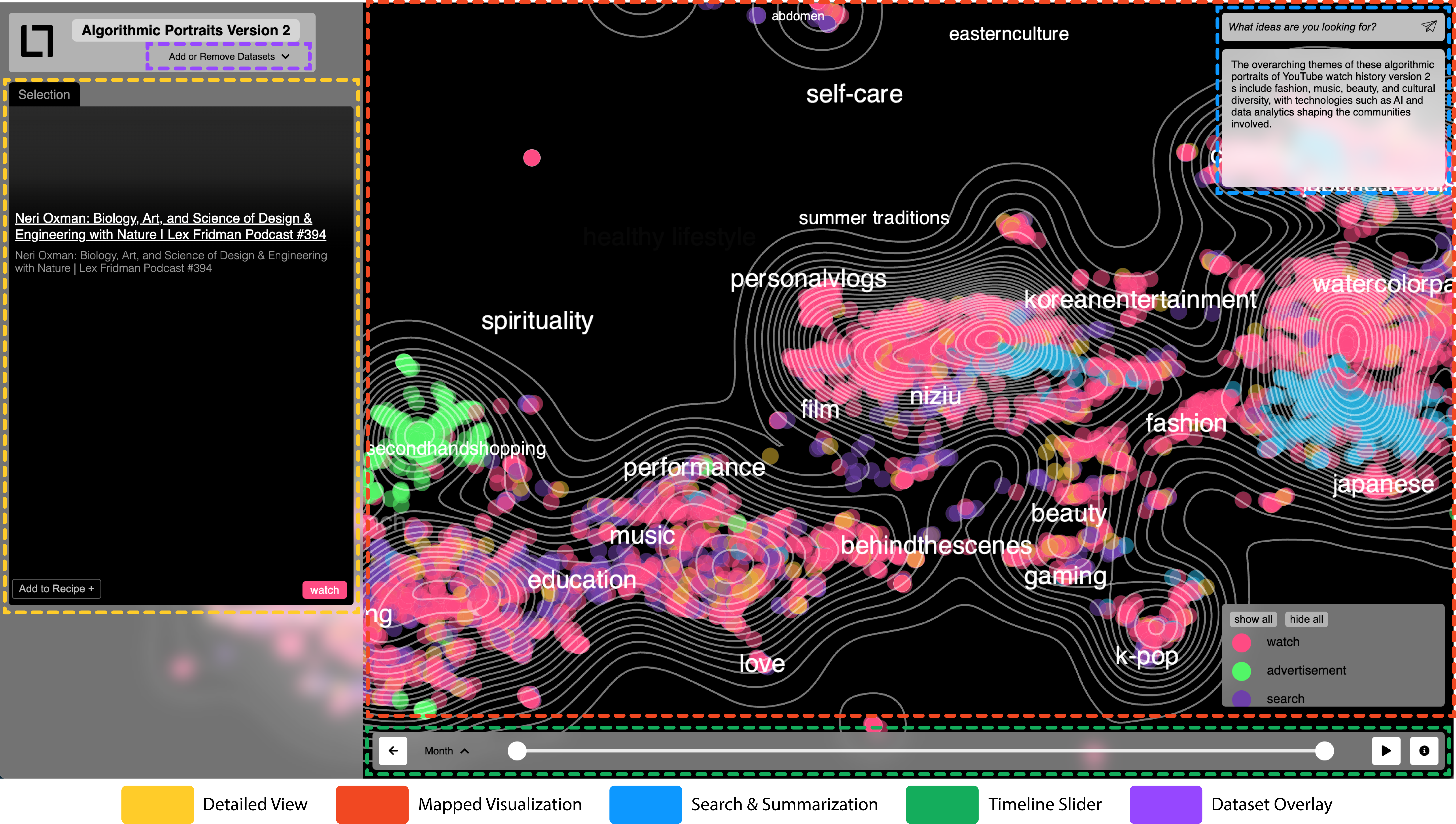}
    \caption{Algorithmic Mirror Interface. We highlight four main components of the interface with dashed lines. The watched videos are in pink, searched terms in purple, watched videos after searching in yellow, watched advertisements in green, and watched shorts videos in blue.}
    \label{fig:algorithmicmirror}
\end{figure*}
Drawing on previous visualization tools to foster awareness and critical reflection of recommendation algorithms, we propose Algorithmic Mirror. This design prototype aims to bring together scattered digital footprints into a single map and visualize how algorithms interpret users' footprints. By doing so, it highlights the influence of filtering algorithms on content exploration and algorithmically-mediated lives.

Algorithmic Mirror is built upon a publicly-available language model-powered knowledge exploration system called Latent Lab~\cite{dunnellLatentLabExploration2023}. The tool provides an automated and easy-to-use pipeline for converting high-dimensional unstructured data into interactive 2D map visualizations. The interface allows users to explore labeled clusters of similar topics and search by semantic context. In our study, participants obtained their YouTube watch history via Google Takeout and uploaded the data to Latent Lab. Below, we discuss the four main key design components of Algorithmic Mirror. 

\textbf{Interactive Map Visualization}
Algorithmic Mirror leverages Latent Lab to create an interactive 2D map of video data, where each dot represents a video. Transcripts are embedded via OpenAI’s text-embedding-ada-002 model and then reduced to two dimensions using UMAP. Clusters signify semantic similarity, and overlaid contour lines illustrate data density. Users can pan, zoom, and view high-level or detailed labels and contours, supported by an occlusion algorithm that prioritizes popular topics for legible labeling.
Clicking any point opens a sidebar showing the detailed view with the video’s header image, title (with a direct link), and its full transcript.

\textbf{Topic Extraction and Data Processing}
Algorithmic Mirror uses Latent Lab to model how recommendation systems analyze digital footprints, inferring user interests and preferences. Latent Lab automatically extracts up to three primary topics from each video transcript using GPT-4o. These topics are aggregated, ranked, and embedded to position them within semantically relevant areas of the map, forming a hierarchical topic tree that scales from broad to specific themes as users zoom. This structure ensures users can navigate clusters with increasing detail and thematic nuance.

\textbf{Summarization and Timeline}
The summarization bar samples transcripts visible on the map, summarizes them via GPT-4o, and presents a concise overview of the current map region. A timeline slider lets users filter videos by date, dynamically updating contour lines and labels to highlight how content shifts over time.

\textbf{Dataset Overlay}
Latent Lab allows users to overlay one dataset onto another, enabling the exploration of one dataset in the context of another by projecting the overlaid points into the semantic space defined by the target dataset's UMAP reducer. This process is non-commutative, meaning overlaying dataset A onto dataset B yields different results than overlaying B onto A, as the spatial relationships depend on the target dataset. In Algorithmic Mirror, this feature is used to analyze cross-platform content, such as overlaying YouTube viewing data onto Spotify listening history (or vice versa) to reveal semantic relationships.
(see Figure \ref{fig:spotify} for visualization of dataset overlay).

\section{User Study}

\textbf{Participants}
We recruited participants through convenience and snowball sampling methods, screening 25 initial respondents to select 14 eligible participants who were 18 or older, visited YouTube daily, and typically browsed from the YouTube homepage. Despite efforts to achieve demographic diversity, our final participant pool showed balanced gender representation but was skewed toward higher education levels, with all participants holding bachelor's degrees, potentially indicating above-average technical and data literacy compared to the general population.

\textbf{Study Procedure}
During a 60-minute user study, each participant was asked to interact with Algorithmic Mirror and answer questions. Each user study contained two parts (see Table \ref{tab:interview-protocol} for details of interview questions). Firstly, in a walk-through and think-aloud of Algorithmic Mirror, participants were allowed to interact freely with Algorithmic Mirror and asked to think aloud, providing insight into their thought processes as they tested the tool~\cite{cookeAssessingConcurrentThinkAloud2010}. Secondly, participants provided feedback on specific design functionalities and were encouraged to critique and reimagine Algorithmic Mirror. All the interviews were conducted in Microsoft Teams due to geographical constraints. The study was approved by the university’s ethics committee. 

\section{Results}
We discuss how Algorithmic Mirror can be used to facilitate critical reflection over platform practices on data collection and data inference.

\textbf{Users developed meta-cognitive awareness of their online behavior.}
After each user uploaded their watched YouTube history (20,000 to 40,000 videos), these records were transformed into topics they could explore. We observed increased engagement after participants started zooming specific interests, eventually clicking on individual dots to look at the details of specific videos. This personalized experience helped the majority of participants ($n=10/14$)  recall content they had watched, including content they did not remember (P5: \textit{``I have gone down like nail, like fingernail polish rabbit hole.''}). This process of recalling and reviewing their content consumption patterns triggered surprise at the breadth and volume of their content consumption across categories (P1: \textit{``It's scary to realize that I'm watching things without awareness.``} P2: \textit{``I'm surprisingly unaware of the gap between my perception and my actual self.''}). 

\textbf{Users discovered fragmented digital identities across platforms.} 
Algorithmic Mirror allows overlay of different datasets, such as comparing YouTube and Spotify consumption patterns. Comparing maps between YouTube-only, Spotify-only, or a combined integrated visualization, P2 realized how \textit{``different personalities exist within each platform.''}, highlighting the fragmented nature of digital self-representation. (see Figure \ref{fig:spotify} for visualization of cross-platform visualization).

\textbf{Users recognized how their interests have evolved over time.}
Most participants ($n=12/14$) interacted with the timeline slider to observe the evolution of their interests over time, often triggering what P5 termed `\textit{nostalgia}'. Participants ($n=8/14$) often first clicked the `Play' button which animates the map month by month, noting that this process allowed them to see what P5 termed ``\textit{watching myself grow}''. P2 stated, \textit{``The tool allows me to think more deeply about the temporal depth and maturity of my interests,''} observing changes in YouTube usage from initially using it as a music tool to later relying more on recommendations. P4 echoed this insight: \textit{``I like the fact that I can see it over time… it does reflect interests that I've had as a person.''} P5 described how \textit{``it takes me back to what I was doing or who I was in the past.''}
In multiple instances, participants then selected specific time ranges by choosing a start and end date in the Timeline Slider. P6 asked: \textit{``What was I looking up when I was taking these exams and when I was on this holiday?''} and used the timeline slider to examine those two moments.
\begin{figure*}[t]
    \includegraphics[width=1.0\linewidth]{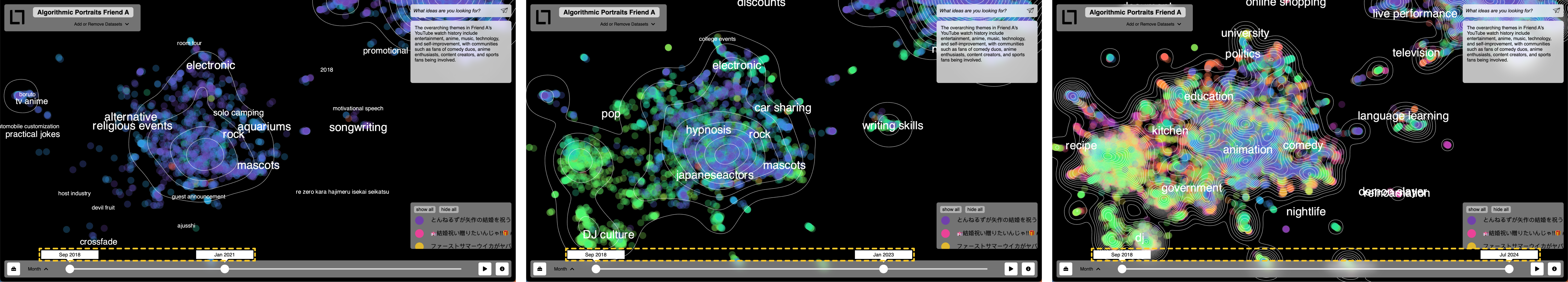}
    \caption{Temporal evolution of Algorithmic Mirror. We animate participant's watch history month by month, starting from September 2018. The first visualization shows data until Jan 2021, the second until Jan 2023, and the third until Jul 2024.}
    \label{fig:temporalevolution}
\end{figure*}

\textbf{Users critically examined inferred data.}
Most participants ($n=9/14$) found that the categories and summaries inferred by Algorithmic Mirror were accurate representations of their interests. They expressed fascination and discomfort with seeing a simulation of how recommendation algorithms on online platforms might interpret their behavior. Looking at the generated summary, P13 noted, ``\textit{The summary looks right. It's always interesting when something interprets you to back yourself}''. This curiosity often led the participants to examine the implications of these algorithmic inferences critically, with P10 noting that the inferred `weight loss' category was for `fitness content` (``\textit{I think watching fitness content means that you can get a lot of ads that have to do with weight loss}''). 
Participants also reflected on the broader effect that platform profiling had on their content consumption. P4 wondered, ``\textit{Has this mirror changed because I changed it or has it changed because the algorithm or some other external thing has changed it?}'' 

\textbf{Users critically examined collected data.}
Looking at maps of inferred interests and generative summaries, some participants ($n=5/14$) questioned why Algorithmic Mirror was so accurate. They experimented with the Timeline Slider to see how much and how long their history had been recorded by YouTube (3 years on average, up to 10 years). P5 observed, ``\textit{They [recommendation algorithms] remember you, like how you were interested.'}' P12 questioned if collecting such data allows platforms to shape future interests, noting, ``\textit{how would it [recommender system] try to track my interest over time and project new interests? [...]  I can imagine it can try to predict how my personality would evolve.'}'

\textbf{Users reflected on alignment between consumption and interests.}
Watching their maps evolve over time, a majority of participants ($n=12/14$) noted that Algorithmic Mirror helped them reflect on whether the viewing consumption aligned with their personal goals and interests. P6 found the tool helpful in identifying content that \textit{``doesn't necessarily spark joy''} but they still continue to watch, prompting them to question engagement with such content. P9 noted, \textit{``The proportion of hobbies I want to pursue […] isn't that high.''} P11 expressed shock, \textit{``I was surprised to see how much time I spent watching things I didn't really want to watch.''} P11 compared the tool to a dietary mirror, saying, 

\begin{quote}
"If you overeat, it shows on your face, and you can look in the mirror. But with YouTube viewing, it's like constantly eating junk food without a mirror to show you've become unhealthy. I probably wasn't aware of it, but now that I can see the mirror, I might decide to cut down on the junk."
\end{quote}

Our findings suggest that algorithmic inference simulations from personal digital footprint enable users to critically evaluate how data collection and inference practices shape their personalized content experiences. 

\section{Design Implications}
This section discusses how design concepts of Algorithmic Mirror address challenges to reflect online behavior on algorithmic feeds and broadly contribute to design development for critical algorithmic literacy.

\subsection{A Single Map for Scattered Digital Footprints}

Assembling scattered digital footprints into a single map, our design mitigates reduced self-awareness~\cite{baughanDonEvenRemember2022} on online platforms. Specifically, the personalized experience can be advanced through two strategies:
\begin{enumerate}[nosep]
    \item Providing both granular and high-level algorithmic categorizations of viewing patterns;
    \item Enabling cross-platform comparison of content consumption behaviors.
\end{enumerate}

First, future designers are encouraged to advance a personalized experience by showing abstract and detailed categories with zooming and panning functions. These functions fostered deeper engagement with categories related to personal experiences they related to most. This design approach goes beyond previous literature on visualizing filter bubbles, which often focuses on specific topics such as news consumption~\cite{gillaniMeMyEcho2018} and provides users with broad categorizations (e.g., ‘immigration’ or ‘market’) ~\cite{chenVisualBubbleExploringHow2022, nagulendraUnderstandingControllingFilter2014}. The ability to see all content was key for participants to compare online consumption with personal interests. 

Second, cross-platform data integration revealed distinct platform-specific behavioral patterns among users. By combining data from YouTube and Spotify, our design enabled users to visualize their complete digital footprint across two platforms. This comprehensive view facilitated novel insights through the analysis of often fragmented data~\cite{donathSocialMachineDesigns2014}, allowing users to critically examine how their interests and behaviors vary across platforms. Cross-platform integration helps users recall and reflect on their watching habits by contextualizing content across different levels of abstraction and multiple platforms. 

\subsection{Hypothetical Inference}

Our design leverages embeddings and language models to simulate how platforms might perceive users' digital footprints, prompting critical reflection by the user on algorithmic inference. While the actual social media algorithms are proprietary, language models can generate plausible inferences from users' extensive historical data (20,000-40,000 data points per user in our study). This advances beyond prior simulations of algorithmic profiling that relied on artificial interactions (e.g. 15--20 videos outside users' typical viewing patterns) on TikTok-like platforms ~\cite{wangCHAITokProofofConceptSystem2024, ailaelkattanAlgorithm2024}. Using years of real watch history enabled more meaningful personal reflection.

The design also improves upon prior approaches by displaying both raw data (e.g., watched videos) and derived inferences (e.g., interest categories) together rather than separately~\cite{nguyenDigitalAgeOur2022,donathDataPortraits2020}. This simultaneous view helped users reflect on how platforms derive inferences from their data, often leading to strong opposition to such practices. Through our design, users gained a critical perspective on recommendation algorithms' societal impact by examining their own data without requiring technical expertise. This personalized lens transformed abstract algorithmic concepts into tangible insights about data collection and inference practices that directly shape users' online experiences. 

\subsection{Considering the Temporal Dimension in Future Design}
Our design utilized long-term data to enable users to reflect on the effects of recommendation algorithms and their actions over time. Algorithmic Mirror invited participants to explore the development of their interests over selected periods using a timeline slider with options ranging from minutes to years. In contrast to static snapshots typical of other approaches ~\cite{gillaniMeMyEcho2018, chenVisualBubbleExploringHow2022, nagulendraUnderstandingControllingFilter2014}, this dynamic approach captures the fluid nature of the algorithmic feeds, which evolves continuously through users’ ongoing engagement.

The temporal visualization also enabled participants to realize the longevity and scale of their digital footprint. Participants expressed surprise at the extent of data retention, which spanned an average of three years and 20,000 data points per user. Future designers could build upon this temporal dimension to foster deeper reflection on the pervasive influence of accumulated data on consumption habits via recommendation systems.

\section{Limitations and Future Work}
Our study focused first on YouTube, a platform providing few options to reflect on content consumption. Using a copy of their data obtained through Google Takeout, users could seamlessly visualize up to 10 years of their watch history. Platforms rarely provide such visually engaging, user-centric presentations of behaviors and preferences. We believe that building tools such as Algorithmic Mirror is key to support reflection on other platforms, including e.g.Facebook, Instagram, or Netflix---but may require additional efforts on data access and design affordances.

In our studies, self-reflection was prompted and encouraged not only by the Algorithmic Mirror tool but also by interacting with the researcher during the user study. One participant doubted they could fully use it without guidance. We believe that research is needed to design such tools with usability and accessibility for a broad audience. Improvements could include introducing mouse-over or pop-up questions to prompt self-reflection without researcher intervention, and presenting more valuable insights in a user-friendly way (such as visualizing time spent rather than the number of videos).

Finally, we showed that Algorithmic Mirror facilitated immediate reflection on content consumption. We hope that future works can build on such efforts to consider how longer-term tools such as Algorithmic Mirror can facilitate not only reflection but also changes in consumption. Participants expressed intent to utilize this tool after our study and potentially change their future behavior regarding online content consumption, yet our study was unable to track further use beyond the experimental period due to time constraints. Longitudinal studies, including participant follow-up, are key. These could involve tracking participants over extended periods to observe behavioral changes 
and identifying specific design elements that promote sustained autonomy over time.

\section{Conclusion}
Our research explored how Algorithmic Mirror could facilitate users' critical reflection over their data on online platforms. Our research offers insights into how individuals can recognize and identify themselves in the digital age to promote more conscious engagement with online content. Our findings suggest design recommendations in XAI field to reassemble scattered digital footprints into a single map, simulate algorithmic inference with LLMs, as well as display a timeline for capturing evolving patterns. We emphasize engaging, user-centric presentations of online behaviors and inferred preferences can enhance critical algorithmic literacy. Future work should aim to reach broader populations with varying educational backgrounds and literacy levels, should promote features that encourage independent reflection, and should follow participants' interactions over a longer period.  

\newpage
\bibliographystyle{ACM-Reference-Format}
\bibliography{reference}

\appendix
\section{Participant Demographics}
\begin{table}[h]
\centering
\begin{tabular}{|c|l|c|c|l|l|}
\hline
ID & Gender & Age Range & Origin & Education & Daily YouTube Session Length\\
\hline
1 & Female & 18-25 & Japanese & Bachelor & < an hour \\
2 & Male & 18-25 & Japanese & Bachelor & < two hours\\
3 & Male & 18-25 & Japanese & Bachelor & > two hours\\
4 & Female & 18-25 & Indian & Master & < an hour \\
5 & Male & 18-25 & Japanese & Bachelor & < an hour \\
6 & Female & 18-25 & Japanese & Bachelor & < an hour \\
7 & Female & 18-25 & Chinese & Bachelor & < two hours \\
8 & Male & 18-25 & Japanese & Bachelor & < two hours \\
9 & Male & 18-25 & Japanese & Bachelor & < an hour \\
10 & Male & 25-30 & American & Master & > two hours \\
11 & Female & 25-30 & American & Master & < two hours \\
12 & Female & 18-25 & American & Master & < an hour \\
13 & Male & 18-25 & Indian & Master & < an hour \\
14 & Male & 25-30 & Japanese & Master & > two hours \\
\hline
\end{tabular}
\caption{Participant Demographics. The sample shows an equal gender distribution (7 males, 7 females), with participants predominantly in the 18-25 age range (11 participants). Most participants are of Japanese origin (8), followed by American (3), Indian (2), and Chinese (1). All participants hold higher education degrees (8 Bachelor's, 6 Master's). YouTube viewing habits indicate most participants (11) watch less than two hours daily, with only 3 participants watching more than two hours.}
\label{tab:demographics}
\end{table}
\newpage
\section{Interview Protocol}
\begin{table}[htbp]
    \centering
    \begin{tabular}{|p{0.95\textwidth}|}
        \hline
        \multicolumn{1}{|m{0.95\textwidth}|}{\centering\vspace{0.5em}\textbf{Interview Protocol: Interaction with Algorithmic Mirror}\vspace{0.5em}} \\
        \hline
        
        \textbf{Phase 1: Initial Exploration} \\[0.3em]
        Please browse this interface freely (10 minutes) \\[1em]
        
        \textbf{Phase 2: User Experience Assessment} \\[0.3em]
        \begin{enumerate}[label=\arabic*., leftmargin=2em, itemsep=0.5em]
            \item What aspects of the tool did you find particularly effective or appealing?
            \item What features or elements did you find challenging or less useful?
            \item Could you suggest specific improvements that would enhance your experience with the tool?
        \end{enumerate} \\[1em]
        
        \textbf{Phase 3: Future Usage Intent} \\[0.3em]
        Would you be interested in continuing to use this interface after the study concludes? \\[0.3em]
        \begin{itemize}[label=$\circ$, leftmargin=2em, itemsep=0.3em]
            \item If yes: What would motivate your continued use?
            \item If no: What factors influence your decision not to continue?
        \end{itemize} \\[1em]
        
        \textbf{Phase 4: Reflection Process Evaluation} \\[0.3em]
        What are your overall thoughts on this reflection process? \\[0.3em]
        \begin{itemize}[label=$\circ$, leftmargin=2em, itemsep=0.3em]
            \item How effective was it in promoting self-reflection?
            \item What impact did it have on your understanding of your online behavior?
        \end{itemize} \\[0.5em]
        \hline
    \end{tabular}
    \caption{Semi-structured Interview Protocol}
    \label{tab:interview-protocol}
\end{table}

\newpage
\section{Dataset Overlay (Spotify and YouTube)}
\begin{figure}[h!]
    \centering
    \includegraphics[width=1\linewidth]{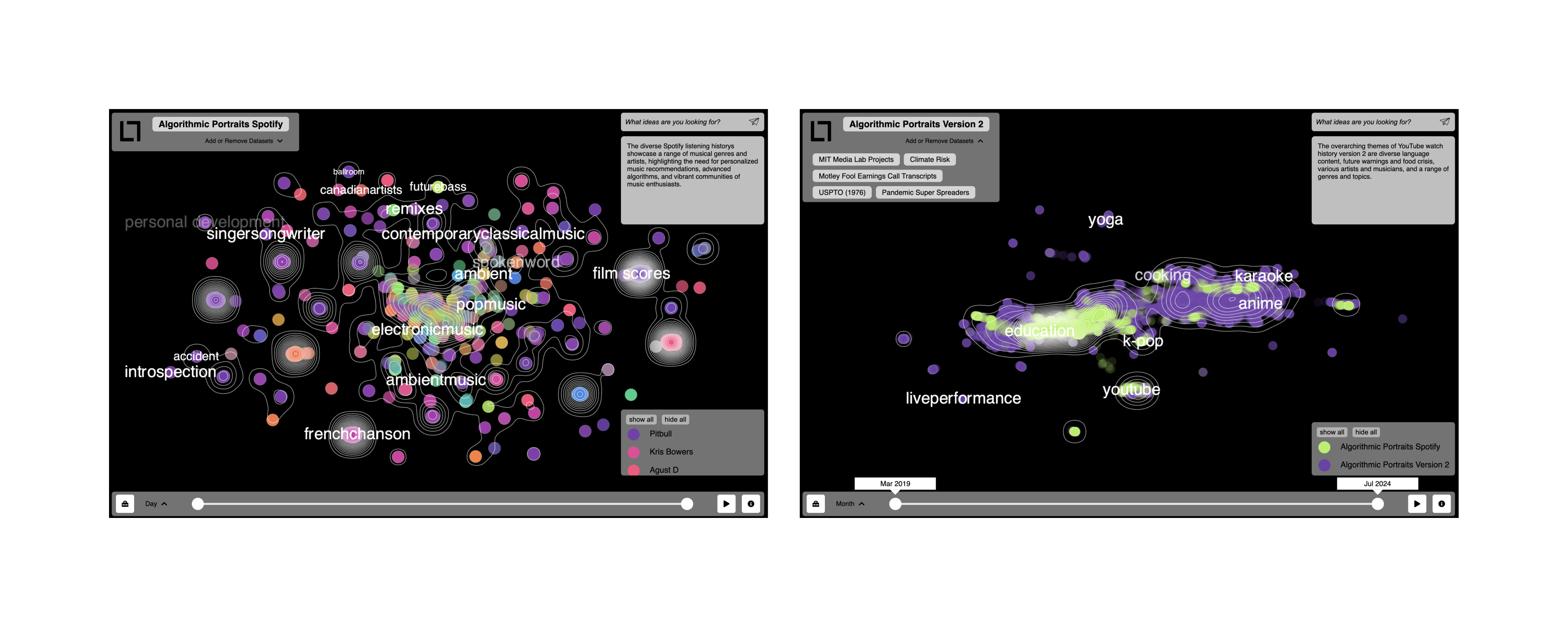}
    \caption{Left panel: Algorithmic Mirror of Spotify listening history, Right panel: Multi-platform Algorithmic Mirror overlaying YouTube viewing data onto Spotify listening history (or vice versa) to reveal semantic relationships. Users can toggle between single-dataset views or combined overlays, with distinct colors (e.g., green for YouTube and purple for Spotify) highlighting cross-platform patterns and thematic connections.}
    \label{fig:spotify}
\end{figure}

\end{document}